\documentclass[letterpaper, 10 pt, conference]{ieeeconf}

\IEEEoverridecommandlockouts 

\usepackage{color}
\usepackage{amsmath,amsthm,amssymb,amsfonts}
\usepackage{graphicx}

\usepackage{amsthm}
\newtheoremstyle{mytheoremstyle} 
    {.3cm}      
    {.3cm}      
    {}          
    {}          
    {\bfseries}  
    {:}         
    {.5em}      
    {}          
\theoremstyle{mytheoremstyle}

\renewcommand{\qedsymbol}{$\blacksquare$}
\newtheorem{thm}{Theorem}

\newtheorem{prop}[thm]{Proposition}

\theoremstyle{definition}

\newtheorem{exmp}{Example}

\newtheorem{assump}{Assumption}

\title{\LARGE \bf
Safety-Critical Control of Compartmental Epidemiological Models \newline with Measurement Delays
}

\author{Tam\'as G. Moln\'ar$^{1}$, Andrew W. Singletary$^{2}$, G\'abor Orosz$^{1,3}$ and Aaron D. Ames$^{2}$
\thanks{$^{1}$Department of Mechanical Engineering, University of Michigan, Ann Arbor, MI 48109, USA {\tt\small molnart@umich.edu, orosz@umich.edu}}%
\thanks{$^{2}$Department of Mechanical and Civil Engineering, California Institute of Technology, Pasadena, CA 91125, USA {\tt\small ames@caltech.edu, asinglet@caltech.edu}}%
\thanks{$^{3}$Department of Civil and Environmental Engineering, University of Michigan, Ann Arbor, MI 48109, USA}%
}

\begin{document}

\maketitle
\thispagestyle{empty}
\pagestyle{empty}

\begin{abstract}

We introduce a methodology to guarantee safety against the spread of infectious diseases by viewing epidemiological models as control systems and by considering human interventions (such as quarantining or social distancing) as control input.
We consider a generalized compartmental model that represents the form of the most popular epidemiological models and we design safety-critical controllers that formally guarantee safe evolution with respect to keeping certain populations of interest under prescribed safe limits.
Furthermore, we discuss how measurement delays originated from incubation period and testing delays affect safety and how delays can be compensated via predictor feedback.
We demonstrate our results by synthesizing active intervention policies that bound the number of infections, hospitalizations and deaths for epidemiological models capturing the spread of COVID-19 in the USA.

\end{abstract}

\section{INTRODUCTION}
\label{sec:intro}

The rapid spreading of COVID-19 across the world forced people to change their lives and practice mitigation efforts at a level never seen before, including social distancing, mask-wearing, quarantining and stay-at-home orders.
These human actions played a key role in reducing the spreading of the virus, although such interventions often have economic consequences, lose of jobs and physiological effects.
Therefore, it is important to focus mitigation efforts and determine when, where and what level of intervention needs to be taken.

This research provides a methodology to determine the level of active human intervention needed to provide safety against the spreading of infection while keeping mitigation efforts minimal.
We use compartmental epidemiological models to describe the spreading of the infection~\cite{giordano2020modelling, kucharski2020early}, and we view these models as control systems where human intervention is the control input.
Viewing epidemiological models as control systems has been proposed in the literature recently~\cite{elhia2013,bolzoni2017,casella2020covid19}, and various models with varying transmission rate~\cite{anderson2020estimating,dandekar2020quantifying,franco2020feedback,weitz2020} have appeared to quantify the level of human interventions in the case of COVID-19.

In this paper, we build on our recent work~\cite{ames2020covid} and use a safety-critical control approach to synthesize control strategies that guide human interventions so that certain safety criteria (such as keeping infection, hospitalization and death below given limits) are fulfilled with minimal mitigation efforts.
The approach is based on the framework of control barrier functions~\cite{ames2014control, ames2016control} that leverages the theory of set invariance~\cite{ames2019control} for dynamical~\cite{brezis1970characterization, abraham2012manifolds} and control systems~\cite{prajna2006barrier, aubin2009viability, blanchini2008set}.
We take into account that data about the spreading of the infection may involve significant measurement delays~\cite{casella2020covid19, Chen2020delays, pei2020initial, Boldog2020} due to the fact that infected individuals may not show symptoms and get tested for quite a few days.
We use predictor feedback control~\cite{Michiels2001, Krstic2010, Karafyllis2017} to compensate these delays, and we provide safety guarantees against errors in delay compensation.

The outline of the paper is as follows.
Section~\ref{sec:model} introduces a generalized compartmental model, which covers the class of the most popular epidemiological models.
Section~\ref{sec:control} introduces safety critical control without considering measurement delays, while Sec.~\ref{sec:delay} is dedicated to delay compensation.
Conclusions are drawn in Sec.~\ref{sec:concl}.


\begin{figure}[t!]
\centering
\includegraphics[width = 0.48\textwidth]{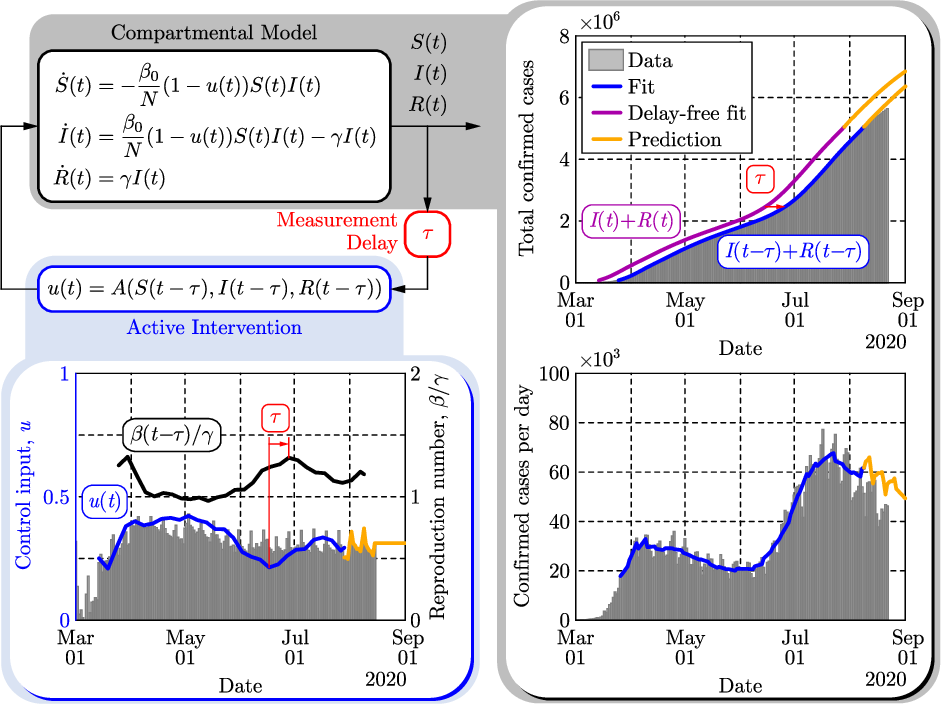}
\caption{
Illustration of the SIR model as control system and its fit to US COVID-19 data~\cite{ames2020covid}.
Model parameters were estimated from compartmental data (right) by accounting for a measurement delay $\tau$.
Mobility data (left) were also used to fit the transmission rate and the associated control input.
}
\label{fig:SIRfit}
\end{figure}

\section{GENERALIZED COMPARTMENTAL MODEL}
\label{sec:model}

Compartmental models describe how the size of certain populations of interest evolve over time.
Consider ${n+m}$ compartments, given by ${x \in \mathbb{R}^{n+m}}$, which are separated into two groups: $n$ so-called multiplicative compartments, given by ${w \in \mathbb{R}^n}$, and $m$ outlet compartments, given by ${z \in \mathbb{R}^m}$.
The evolution of these compartments over time $t$ can be given by the following {\em generalized compartmental model}:
\begin{align}
\begin{split}
\dot{w}(t) & = f(w(t)) + g(w(t)) u(t), \\
\dot{z}(t) & = q(w(t)) + r(z(t)),
\end{split}
\label{eq:model}
\end{align}
where ${x = [w^{\rm T}\ z^{\rm T}]^{\rm T}}$, initial conditions are ${x(0) = x_{0}}$, while ${f,g: \mathbb{R}^{n} \to \mathbb{R}^{n}}$, ${q: \mathbb{R}^{n} \to \mathbb{R}^{m}}$ and ${r: \mathbb{R}^{m} \to \mathbb{R}^{m}}$ are assumed locally Lipschitz continuous and
depend on the choice of the model; see Examples~\ref{ex:SIR},~\ref{ex:SEIR} and~\ref{ex:SIHRD}.

In~(\ref{eq:model}), the multiplicative compartments $w$ are populations that essentially describe the transmission of the infection.
The transmission can be reduced by active interventions, whose intensity is quantified by a control input ${u \in \mathcal{U} \subset \mathbb{R}}$  (considered here as scalar, although multiple inputs could be studied analogously).
The outlet compartments $z$ do not actively govern transmission, but rather indicate its effects, as their evolution is driven by the multiplicative compartments.


\begin{exmp} \label{ex:SIR}
\textbf{SIR model.}
One of the most fundamental epidemiological models is the {\em SIR model}~\cite{hethcote2000mathematics, batista2020estimation}
that consists of {\em susceptible}, $S$, {\em infected}, $I$, and {\em recovered}, $R$, populations.
The SIR model captures the spread of the infection based on the interplay between the susceptible and infected populations.
Thus, $S$ and $I$ are multiplicative compartments, while $R$, that measures the number of recovered (or deceased) individuals, is an outlet compartment.
The model uses three parameters: the transmission rate ${\beta_0>0}$, the recovery rate ${\gamma>0}$ and the total population $N$.
Active interventions given by the control input ${u \in [0,1]}$ allow the population to reduce the transmission to an effective rate ${\beta=\beta_0(1-u)}$, where ${u=0}$ means no intervention and ${u=1}$ means total isolation of infected individuals.
This puts the SIR model with active intervention to form~(\ref{eq:model}) where
\begin{alignat}{6}
w & =
\begin{bmatrix}
S \\ I
\end{bmatrix}\!, \quad
&& f(w) && =
\begin{bmatrix}
-\frac{\beta_{0}}{N} S I \\
\frac{\beta_{0}}{N} S I - \gamma I
\end{bmatrix}\!, \quad
&& g(w) && =
\begin{bmatrix}
\frac{\beta_{0}}{N} S I \\
-\frac{\beta_{0}}{N} S I
\end{bmatrix}\!, \nonumber \\
z & = R,
&& q(w) && = \gamma I,
&& r(z) && = 0.
\label{eq:SIR}
\end{alignat}

\end{exmp}

\begin{exmp} \label{ex:SEIR}
\textbf{SEIR model.} The {\em SEIR model}~\cite{Blyuss2020SEIR, lopez2020modifiedSEIR} is an extension of the SIR model that incorporates an {\em exposed} population $E$ apart from the $S$, $I$ and $R$ compartments.
The exposed individuals are infected but not yet infectious over a latency period given by ${1/\sigma > 0}$.
Since the latency affects the transmission, $E$ is a multiplicative compartment.
The SEIR model can be described by~(\ref{eq:model}) with
\begin{alignat}{6}
& w =
\begin{bmatrix}
S \\ E \\ I
\end{bmatrix}\!, \quad
&& f(w) && =
\begin{bmatrix}
-\frac{\beta_{0}}{N} S I \\
\frac{\beta_{0}}{N} S I - \sigma E \\
\sigma E - \gamma I
\end{bmatrix}\!, \quad
&& g(w) && =
\begin{bmatrix}
\frac{\beta_{0}}{N} S I \\
-\frac{\beta_{0}}{N} S I \\
0
\end{bmatrix}\!, \nonumber \\
& z = R,
&& q(w) && = \gamma I,
&& r(z) && = 0.
\label{eq:SEIR}
\end{alignat}

\end{exmp}

\begin{exmp} \label{ex:SIHRD}
\textbf{SIHRD model.} The {\em SIHRD model}~\cite{ames2020covid} adds two more outlet compartments to the SIR model: hospitalized population $H$ and deceased population $D$.
Their evolution is captured by three additional parameters: the hospitalization rate ${\lambda>0}$, the recovery rate ${\nu>0}$ in hospitals and the death rate ${\mu>0}$.
Equation~(\ref{eq:model}) yields the SIHRD model for
\begin{alignat}{6}
& w \!=\!
\begin{bmatrix}
S \\ I
\end{bmatrix}\!, \;\;
&& f(w) && \!=\!
\begin{bmatrix}
-\frac{\beta_{0}}{N} S I \\
\frac{\beta_{0}}{N} S I \!-\! (\gamma \!+\! \lambda \!+\! \mu) I
\end{bmatrix}\!, \;\;
&& g(w) && \!=\!
\begin{bmatrix}
\frac{\beta_{0}}{N} S I \\
-\frac{\beta_{0}}{N} S I
\end{bmatrix}\!, \nonumber \\
& z \!\!=\!\!
\begin{bmatrix}
H \\ R \\ D
\end{bmatrix}\!,
&& q(w) && \!=\!
\begin{bmatrix}
\lambda I \\ \gamma I \\ \mu I
\end{bmatrix}\!,
&& r(z) && \!=\!
\begin{bmatrix}
-\nu H \\ \nu H \\ 0
\end{bmatrix}\!.
\label{eq:SIHRD}
\end{alignat}

\end{exmp}

There exist several other compartmental models of form~(\ref{eq:model}) which involve further compartments, such as the SIRD~\cite{fernandez2020estimating}, SIRT~\cite{dandekar2020quantifying}, SIXRD~\cite{Humphries2020} or SIDARTHE~\cite{giordano2020modelling} models.
More complex models can provide higher fidelity, although they involve more parameters that need to be identified.
In what follows, we show applications of the SIR and SIHRD models and we discuss the occurrence of time delays related to incubation and testing.
We omit further discussions on latency, the SEIR model or other more complex models.

Fig.~\ref{fig:SIRfit} shows the performance of the SIR model in capturing the spread of COVID-19 for the case of US national data.
The parameters ${\beta_0=0.33\,{\rm day^{-1}}}$, ${\gamma=0.2\,{\rm day^{-1}}}$ and ${N=33 \times 10^6}$ of the SIR model and the control input $u(t)$ were fitted following the algorithm in~\cite{ames2020covid} to the recorded number of confirmed cases ${I+R}$~\cite{coviddata} between March 25 and August 9, 2020 and to mobility data~\cite{mobilitydata} about the medium time people spent home.
The fitted control input (blue) follows the trend of the mobility data (gray) well in the beginning of the pandemic when stay-at-home orders came into action, and it deviates later when other means of mitigation (such as mask-wearing) became more significant.
While the fitted model (blue) captures the data about confirmed cases (gray), the model also has predictive power (orange); see more details about forecasting in~\cite{ames2020covid}.

Note that once an individual gets infected by COVID-19, it takes a few days of incubation period to show symptoms and an additional few days to get tested for the virus~\cite{casella2020covid19, Chen2020delays, pei2020initial, Boldog2020}.
Therefore, the measured number of confirmed cases represents a delayed state of the system, ${I(t-\tau)+R(t-\tau)}$, and thus we involved a time delay $\tau$ in the model identification process, which was found to be ${\tau=11\,{\rm days}}$ by fitting~\cite{ames2020covid}.
The delay-free counterpart of the fit (purple) shows that the measurement delay can lead to a significant error in identifying the true current level of infection.
The effects of the delay $\tau$ on safety-critical control and its compensation will be discussed in Sec.~\ref{sec:delay}.


\section{SAFETY-CRITICAL CONTROL}
\label{sec:control}

Formally, safety can be translated into keeping system~(\ref{eq:model}) within a {\em safe set} ${\mathcal{S} \subset \mathbb{R}^{n+m}}$ that is the 0-superlevel set of a continuously differentiable function ${h: \mathbb{R}^{n+m} \to \mathbb{R}}$:
\begin{equation}
\mathcal{S} := \{ x  \in \mathbb{R}^{n+m} ~ : ~ h(x) \geq 0 \},
\label{eq:safeset}
\end{equation}
where ${x = [w^{\rm T},\ z^{\rm T}]^{\rm T}}$.
Function $h$ prescribes the condition for safety: for example, if one intends to keep the infected population $I$ under a limit $I_{\rm max}$ for the SIR, SEIR or SIHRD models, the safety condition is ${h(x)=I_{\rm max} - I \geq 0}$.

To guarantee safety, we design a locally Lipschitz continuous controller
\begin{equation}
u(t) = A (x(t))
\label{eq:controller}
\end{equation}
that ensures that the set $\mathcal{S}$ in~(\ref{eq:safeset}) is forward invariant under the dynamics~(\ref{eq:model}), i.e., if ${x(0) \in \mathcal{S}}$ (${h(x(0)) \geq 0}$), then ${x(t) \in \mathcal{S}}$ (${h(x(t)) \geq 0}$) for all ${t>0}$. 
Below we use the framework of {\em control barrier functions}~\cite{ames2014control, ames2016control} to synthesize controllers that are able to keep certain compartments of interest within prescribed limits.
First, we consider safety for multiplicative compartments, and then for outlet compartments.

\subsection{Safety Guarantees for Multiplicative Compartments}
\label{sec:multiplicative}

Consider keeping the $i$-th multiplicative compartment (${1 \leq i \leq n}$) below a safe limit given by $C_{i}$, i.e., we prescribe
\begin{equation}
h(x) = C_{i} - w_{i},
\label{eq:h_multi}
\end{equation}
where $C_{i}$ is an upper bound for $w_{i}$.
A lower bound could also be considered similarly, by taking ${h(x)=w_{i} - C_{i}}$.

\begin{thm}
Consider dynamical system~(\ref{eq:model}), function $h$ in~(\ref{eq:h_multi}) and 
the corresponding set $\mathcal{S}$ given by~(\ref{eq:safeset}).
The following safety-critical active intervention controller guarantees that $\mathcal{S}$ is forward invariant (safe) under dynamics~(\ref{eq:model}) if ${g_{i}(w) \neq 0}$, ${\forall w \in \mathbb{R}^n}$:
\begin{equation}
u(t) \!=\! A_{i}(x(t)) \!=\! - {\rm sign}(g_{i}(w(t))) \mathrm{ReLU}\! \left(\! \frac{\varphi_{i}(w(t))}{|g_{i}(w(t))|} \!\right)\!,
\label{eq:controller_multi_ReLU}
\end{equation}
where ${\mathrm{ReLU}(\cdot) = \max \{0, \cdot \}}$ is the rectified linear unit,
\begin{equation}
\varphi_{i}(w) = f_{i}(w) - \alpha(C_{i} - w_{i})
\label{eq:phi_multi}
\end{equation}
and ${\alpha > 0}$.
Furthermore, the controller is optimal in the sense that it has minimum-norm control input.
\end{thm}

\proof
According to~\cite{ames2016control}, the necessary and sufficient condition of forward set invariance is given by\footnote{More precisely, $\alpha$ must be chosen as an extended class $\mathcal{K}$ function~\cite{ames2016control}, but we use a constant for simpler discussion and without loss of generality.}
\begin{equation}
\dot{h}(x(t))
\geq - \alpha h(x(t)),
\label{eq:safetycondition}
\end{equation}
${\forall t \geq 0}$, where the derivative is taken along the solution of~(\ref{eq:model}).
If there exists a control input $u(t)$ so that~(\ref{eq:safetycondition}) is satisfied, then $h$ is called a {\em control barrier function}.
Substitution of~(\ref{eq:h_multi}) and~(\ref{eq:model}) into~(\ref{eq:safetycondition}) gives the safety condition
\begin{equation}
- \varphi_{i}(w(t)) - g_{i}(w(t)) u(t) \geq 0,
\label{eq:safetycondition_multi}
\end{equation}
where $\varphi_{i}$ is given by~(\ref{eq:phi_multi}).
The control input $u(t)$ must satisfy~(\ref{eq:safetycondition_multi}) for all ${t \geq 0}$.
To keep control efforts minimal, one can achieve this by solving the quadratic program:
\begin{align}
\begin{split}
u(t) = A_{i}(x(t)) = \mathrm{arg}\hspace{-.1cm}\min_{\hspace{-.2cm} u \in \mathcal{U}} & \quad u^2  \\
\mathrm{s.t.} & \quad (\ref{eq:safetycondition_multi}).
\end{split}
\label{eq:QP_multi}
\end{align}
Based on the KKT conditions~\cite{Boyd2004}, the explicit solution is
\begin{align}
u(t) \!=\! A_{i}(x(t)) \!=\!
\begin{cases}
0 & \mathrm{if} \;\; - \varphi_{i}(w(t)) \geq 0, \\
-\frac{\varphi_{i}(w(t))}{g_{i}(w(t))} & \mathrm{if} \;\; - \varphi_{i}(w(t)) < 0,
\end{cases}
\label{eq:controller_multi}
\end{align}
if ${g_{i}(w(t)) \neq 0}$, which can be simplified to~(\ref{eq:controller_multi_ReLU}).
\hfill \qedsymbol

We remark that if ${g_{i}(w) = 0}$, safety can be ensured by the help of {\em extended control barrier functions} as discussed for the safety guarantees of outlet compartments in Sec.~\ref{sec:outlet}.




\begin{figure}[t!]
\centering
\includegraphics[width = 0.48\textwidth]{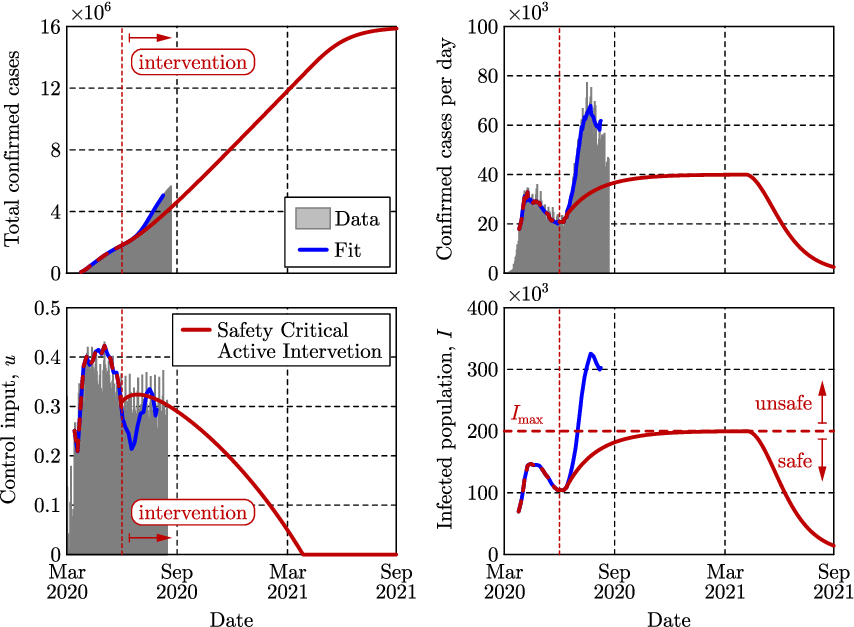}
\caption{
Safety-critical active intervention control of the SIR model fitted in Fig.~\ref{fig:SIRfit} to US COVID-19 data.
The controller keeps the infected population under the prescribed limit $I_{\rm max}$ as opposed to the second wave of infection experienced over the summer of 2020 due to a drop in mitigation efforts.
}
\label{fig:SIRcontrol}
\end{figure}

For example, to keep the infected population $I$ below the limit $I_{\rm max}$ for the SIR model given by~(\ref{eq:SIR}), one shall prescribe ${h(x)=I_{\rm max} - I}$, and~(\ref{eq:controller_multi_ReLU}) leads to the controller
\begin{equation}
A_{I}(x) = \mathrm{ReLU} \left( 1 - \frac{\alpha_{I} (I_\mathrm{max} - I) + \gamma I}{\beta_{0} S I / N} \right).
\label{eq:AI}
\end{equation}
Fig.~\ref{fig:SIRcontrol} shows the closed loop dynamics of the COVID-19 model of Fig.~\ref{fig:SIRfit} by prescribing ${I_\mathrm{max} = 200,000}$ and using ${\alpha=\gamma/10}$.
Indeed, the safety-critical controller (red) applied from June 1, 2020 (with the blue fitted control input as history) keeps the level of infection under the safe limit (red dashed), while gradually reducing mitigation efforts to zero.
Meanwhile, the US experienced a second wave of infections (gray) in the summer of 2020, which was caused by the drop in mitigation efforts in June (see the blue control input).

\subsection{Safety Guarantees for Outlet Compartments}
\label{sec:outlet}

Now consider the case where the $j$-th outlet compartment (${1 \leq j \leq m}$) needs to be kept within the safe limit $C_{j}$:
\begin{equation}
h(x) = C_{j} - z_{j}.
\label{eq:h_outlet}
\end{equation}
In the following theorem, we use a dynamic extension of control barrier functions to guarantee safety.

\begin{thm}
Consider dynamical system~(\ref{eq:model}), function $h$ in~(\ref{eq:h_outlet}) and
the corresponding set $\mathcal{S}$ given by~(\ref{eq:safeset}).
The following safety-critical active intervention controller guarantees that $\mathcal{S}$ is forward invariant (safe) under dynamics~(\ref{eq:model}) if ${\dot{h}(x(0)) + \alpha h(x(0)) \geq 0}$ and if ${L_{g}q_{j}(w) \neq 0}$, ${\forall w \in \mathbb{R}^n}$:
\begin{equation}
u(t) \!=\! A_{j}(x(t)) \!=\! - {\rm sign}\! \left( L_{g}q_{j}(w(t)) \right) \! \mathrm{ReLU} \! \left(\! \frac{\varphi_{j}^{\rm e}(x(t))}{\left| L_{g}q_{j}(w(t)) \right|} \!\right)\!,
\label{eq:controller_outlet_ReLU}
\end{equation}
where ${L_{g}q_{j}(w) = {\frac{\partial q_{j}}{\partial w}(w) g(w)}}$,
\begin{multline}
\varphi_{j}^{\rm e}(x) = \frac{\partial q_{j}}{\partial w}(w) f(w)
+ \frac{\partial r_{j}}{\partial z}(z) \big( q(w) + r(z) \big) \\
+ (\alpha + \alpha^{\rm e}) \big( q_j(w) + r_{j}(z) \big)
- \alpha^{\rm e} \alpha (C_{j} - z_{j}),
\label{eq:phi_outlet}
\end{multline}
and ${\alpha>0}$, ${\alpha^{\rm e}>0}$.
Furthermore, the controller is optimal in the sense that it has minimum-norm control input.
\end{thm}

\proof
We again use~(\ref{eq:safetycondition}) as the necessary and sufficient condition for safety, where the following expression appears:
\begin{multline}
h^{\rm e}(x(t)) :=  \dot{h}(x(t)) + \alpha h(x(t)) \\
= - (q_{j}(w(t)) + r_{j}(z(t))) + \alpha (C_{j} - z_{j}(t)),
\label{eq:extendedbarrier}
\end{multline}
which puts the safety condition into the form ${h^{\rm e}(x(t)) \geq 0}$, ${\forall t \geq 0}$.
However, the control input does not explicitly show up in~(\ref{eq:extendedbarrier}).
Still, if there exists a control input that satisfies
\begin{equation}
\dot{h}^{\rm e}(x(t))
\geq - \alpha^{\rm e} h^{\rm e}(x(t)),
\label{eq:safetycondition_extended}
\end{equation}
then $h^{\rm e}$ is an {\em extended control barrier function}~\cite{nguyen2016exponential, ames2019control}, whose 0-superlevel is forward invariant, that is, ${h^{\rm e}(x_{0}) \geq 0}$ implies ${h^{\rm e}(x(t)) \geq 0}$, ${\forall t > 0}$.
Substitution of~(\ref{eq:extendedbarrier}),~(\ref{eq:h_outlet}) and~(\ref{eq:model}) into~(\ref{eq:safetycondition_extended}) gives the extended safety condition
\begin{equation}
- \varphi_{j}^{\rm e}(x(t)) - L_{g}q_{j}(w(t)) u(t) \geq 0,
\label{eq:safetycondition_outlet}
\end{equation}
where $\varphi_{j}^{\rm e}$ is defined by~(\ref{eq:phi_outlet}).
This can be satisfied by a min-norm controller obtained from the quadratic program:
\begin{align}
\begin{split}
u(t) = A_{j}(x(t)) = \mathrm{arg}\hspace{-.1cm}\min_{\hspace{-.2cm} u \in \mathcal{U}} & \quad u^2  \\
\mathrm{s.t.} & \quad (\ref{eq:safetycondition_outlet}).
\end{split}
\label{eq:QP_outlet}
\end{align}
The explicit solution of the quadratic program is
\begin{align}
u(t) \!=\! A_{j}(x(t)) \!=\!
\begin{cases}
0 & \mathrm{if} \;\, - \varphi_{j}^{\rm e}(x(t)) \!\geq\! 0, \\
-\frac{\varphi_{j}^{\rm e}(x(t))}{L_{g}q_{j}(w(t))} & \mathrm{if} \;\, - \varphi_{j}^{\rm e}(x(t)) \!<\! 0,
\end{cases}
\label{eq:controller_outlet}
\end{align}
if ${L_{g}q_{j}(w(t)) \neq 0}$, which is equivalent to~(\ref{eq:controller_outlet_ReLU}).
\hfill \qedsymbol



As an example of keeping outlet compartments safe, consider limiting the number of hospitalizations below $H_{\rm max}$ and deaths below $D_{\rm max}$ for the SIHRD model given by~(\ref{eq:SIHRD}).
By choosing ${h(x)=H_{\rm max} - H}$, one can guarantee safety in terms of hospitalization based on~(\ref{eq:controller_outlet_ReLU}) by the controller:
\begin{multline}
A_{H}(x)
= \mathrm{ReLU} \left( 1  - \frac{\alpha_{H}^{\rm e} \alpha_{H} (H_\mathrm{max} - H)}{\lambda \beta_{0} S I / N} \right. \\
\left. - \frac{(\nu - \alpha_{H} - \alpha_{H}^{\rm e}) (\lambda I - \nu H) + (\gamma + \lambda \!+\! \mu) \lambda I}{\lambda \beta_{0} S I / N} \right),
\label{eq:AH}
\end{multline}
whereas prescribing ${h(x) = D_{\rm max} - D}$ ensures safety by upper bounding deaths via:
\begin{multline}
A_{D}(x) = \mathrm{ReLU} \left( 1 - \frac{\alpha_{D}^{\rm e} \alpha_{D} (D_\mathrm{max}-D)}{\mu \beta_{0} S I / N} \right. \\
\left. -\frac{(\gamma+\lambda+\mu-\alpha_{D}-\alpha_{D}^{\rm e}) \mu I}{\mu \beta_{0} S I / N} \right).
\label{eq:AD}
\end{multline}

\subsection{Safety Guarantees for a Combination of Compartments}
\label{sec:multiplebarrier}

Having synthesized controllers that keep selected compartments safe, let us now guarantee safety for multiple compartments at the same time: a set of multiplicative compartments ${\mathcal{I} \subset \{1, \ldots, n\}}$ and a set of outlet compartments ${\mathcal{J} \subset \{1, \ldots, m\}}$.
To formulate the safety condition, one can utilize~(\ref{eq:safetycondition_multi}) for any multiplicative compartment ${i \in \mathcal{I}}$ and~(\ref{eq:safetycondition_outlet}) for any outlet compartment ${j \in \mathcal{J}}$.
Then, one needs to solve the corresponding quadratic program subject to all these constraints.
In general, the quadratic program can only be solved numerically and one may need relaxation terms to satisfy multiple constraints~\cite{ames2016control}.
However, analytical solutions can be found in some special cases, such as the one given by the following assumption.

\begin{assump}
\label{assump:sign}
Assume that the following terms have the same sign:
${{\rm sign}(g_{i}(w(t)))={\rm sign}(L_{g}q_{j}(w(t))) =-1}$, ${\forall i \in \mathcal{I}}$, ${\forall j \in \mathcal{J}}$, ${\forall t \geq 0}$.
\end{assump}

This assumption often holds for models where compartments need to be upper bounded for safety, e.g., the assumption holds for keeping $E$, $I$, $R$, $H$ or $D$ below a safe limit in the SIR, SEIR or SIHRD models.
Under this assumption, one can state the following proposition.

\begin{prop}
Consider dynamical system~(\ref{eq:model}) with Assumption~\ref{assump:sign} and the controllers~(\ref{eq:controller_multi_ReLU}) and~(\ref{eq:controller_outlet_ReLU}) that keep individual multiplicative compartments $w_{i}$, ${i \in \mathcal{I} \subset \{ 1, \ldots, n \}}$ and outlet compartments $z_{j}$, ${j \in \mathcal{J} \subset \{ 1, \ldots, m \}}$ safe using the control barrier functions in~(\ref{eq:h_multi}) and~(\ref{eq:h_outlet}).
The following safety-critical active intervention controller guarantees safety for all compartments at the same time:
\begin{equation}
u(t) = \max_{i \in \mathcal{I}, j \in \mathcal{J}} \left\{ A_{i}(x(t)), A_{j}(x(t)) \right\}.
\label{eq:controller_combined}
\end{equation}
That is, one needs to take the maximum of the individual control inputs that keep each individual compartment safe.

\end{prop}

\proof
If Assumption~\ref{assump:sign} holds, the safety conditions in~(\ref{eq:safetycondition_multi}) and~(\ref{eq:safetycondition_outlet}) can be combined into one inequality:
\begin{equation}
\min_{i \in \mathcal{I}, j \in \mathcal{J}} \left\{
\frac{- \varphi_{i}(w(t))}{|g_{i}(w(t))|}, 
\frac{- \varphi_{j}^{\rm e}(x(t))}{| L_{g}q_{j}(w(t)) |}
\right\} + u(t) \geq 0.
\label{eq:safetycondition_combined}
\end{equation}
Then, one can solve the quadratic program:
\begin{align}
\begin{split}
u(t) = A_{\mathcal{I}\mathcal{J}}(x(t)) = \mathrm{arg}\hspace{-.1cm}\min_{\hspace{-.2cm} u \in \mathcal{U}} & \quad u^2  \\
\mathrm{s.t.} & \quad (\ref{eq:safetycondition_combined})
\end{split}
\label{eq:QP_combined}
\end{align}
in the form:
\begin{equation}
u(t) \!=\! \mathrm{ReLU} \! \left(\! -\!\!\!\min_{i \in \mathcal{I}, j \in \mathcal{J}} \!\!\left\{\!
\frac{- \varphi_{i}(w(t))}{|g_{i}(w(t))|}, 
\frac{- \varphi_{j}^{\rm e}(x(t))}{| L_{g}q_{j}(w(t)) |}
\!\right\} \!\right)\!.
\end{equation}
This can be simplified to~(\ref{eq:controller_combined}) based on~(\ref{eq:controller_multi_ReLU}) and~(\ref{eq:controller_outlet_ReLU}).
\hfill \qedsymbol

\begin{figure}[t!]
\centering
\includegraphics[width = 0.48 \textwidth]{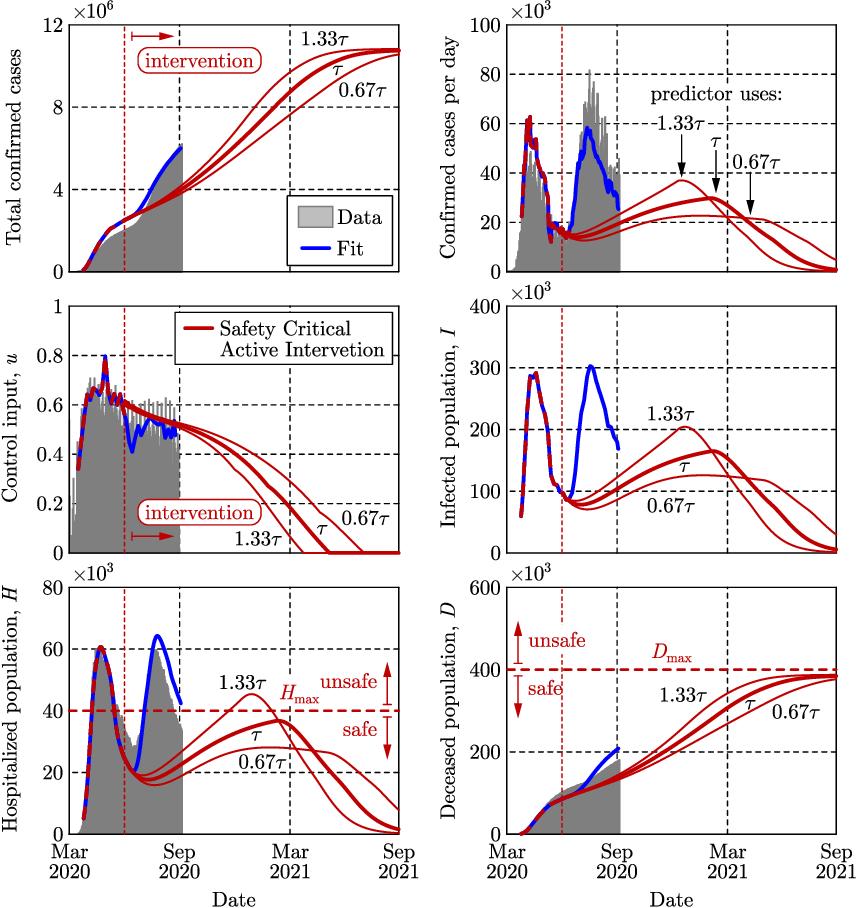}
\caption{
The dynamics of the SIHRD model fitted to US COVID-19 data under the safety-critical active intervention policy that keeps hospitalization and deaths under the prescribed limits $H_\mathrm{max}$ and $D_\mathrm{max}$.
}
\label{fig:SIHRDcontrol}
\end{figure}

Fig.~\ref{fig:SIHRDcontrol} shows the closed loop response of the SIHRD model given by~(\ref{eq:SIHRD}) that was fitted to US COVID-19 data~\cite{ames2020covid}.
The data about confirmed cases were scaled by the cube root of the positivity rate (positive per total tests) to account for the significant under-reporting of cases during the first wave of the virus (and cube root was applied to scale less aggressively).
Starting from June 1, safety-critical active intervention control is applied to limit both the hospitalizations below ${H_{\rm max} = 40,000}$ and the deaths below ${D_{\rm max} = 400,000}$.
Based on~(\ref{eq:controller_combined}), we utilize the controller ${A_{HD}(x)=\max\{A_{H}(x),A_{D}(x)\}}$ where ${A_{H}}$ and $A_{D}$ are given by~(\ref{eq:AH}) and~(\ref{eq:AD}).
The model and controller parameters are
${\beta_0=0.53\,{\rm day^{-1}}}$,
${\gamma=0.14\,{\rm day^{-1}}}$,
${\lambda=0.03\,{\rm day^{-1}}}$,
${\nu\!=\!0.14\,{\rm day^{-1}}}$,
${\mu\!=\!0.01\,{\rm day^{-1}}}$,
${N\!=\!15 \!\times\! 10^6}$,
${\tau\!=\!9\,{\rm days}}$,
${\alpha_{D}\!=\!\alpha_{D}^{\rm e}=\alpha_{H}=(\gamma+\lambda+\mu)/10}$
and
${\alpha_{H}^{\rm e}\!=\!\nu/10}$.
Safety-critical control is able to reduce mitigation efforts
while keeping the system below the prescribed hospitalization and death bounds and preventing a second wave of the virus.

\section{SAFETY UNDER MEASUREMENT DELAYS}
\label{sec:delay}

Controller~(\ref{eq:controller}) in Sec.~\ref{sec:control} is designed based on feeding back the {\em instantaneous state} $x(t)$ of the compartmental model.
However, data about certain compartments is measured with delay due to the incubation period and testing delays.
Thus, the instantaneous state $x(t)$ may not be available for feedback, but the {\em delayed state} ${x(t-\tau)}$ with measurement delay $\tau$ shall be used.
If one implements ${A(x(t-\tau))}$ instead of $A(x(t))$ for active intervention, a significant discrepancy between the delayed and instantaneous states can endanger safety.
For example, the delay was identified to be ${\tau=11\,{\rm days}}$ for the US COVID-19 data in Fig.~\ref{fig:SIRfit}, while the infected population grew from a few thousands to more than a hundred thousand within 11 days in mid March.
This difference significantly impacts safety-critical control.
Thus, we propose a method to compensate delays via predicting the instantaneous state from the delayed one and we analyze how the prediction error affects safety.








\subsection{Safety of Predictor Feedback Control}

We use the idea of {\em predictor feedback control}~\cite{Michiels2001, Krstic2010, Karafyllis2017}
to overcome the effect of delays.
At each time moment $t$ we use the data that are available up to time ${t-\tau}$ and we calculate a {\em predicted state}
$x_{\rm p}(t)$ that approximates the instantaneous state: ${x_{\rm p}(t) \approx x(t)}$.
Then, we use the predicted state in the feedback law by applying
${A(x_{\rm p}(t)) \approx A(x(t))}$.
If the prediction is perfect
(i.e., ${x_{\rm p}(t) = x(t)}$),
safety is guaranteed even in the presence of delay according to Sec.~\ref{sec:control}.
Below we analyze how errors in the prediction affect safety.

The prediction can be done by any model-based or data-based methods; see Example~\ref{exmp:pred} for instance.
At this point we only assume that the prediction error defined by
\begin{equation}
e(t) := x_{\rm p}(t) - x(t)
\label{eq:predictionerror}
\end{equation}
is bounded in the sense that
${\| e(t) \|_{\infty} \leq \varepsilon}$
for some ${\varepsilon \geq 0}$.
The prediction error leads to an input disturbance \begin{equation}
d(t) := A(x_{\rm p}(t)) - A(x(t))
\label{eq:disturbance}
\end{equation}
relative to the nominal control input ${u(t) = A(x(t))}$, which yields the closed control loop
\begin{align}
\begin{split}
\dot{w}(t) & = f(w(t)) + g(w(t)) (u(t)+d(t)), \\
\dot{z}(t) & = q(w(t)) + r(z(t)).
\end{split}
\label{eq:closedloop_disturbance}
\end{align}
For a Lipschitz continuous controller $A$ (such as~(\ref{eq:controller_multi_ReLU}) or~(\ref{eq:controller_outlet_ReLU})) with Lipschitz constant $c$, the disturbance is upper bounded by ${\| d(t) \|_{\infty} \leq c \| e(t) \|_{\infty} \leq c \varepsilon =: \delta}$.
The following theorem summarizes how the disturbance affects safety via the notion of {\em input-to-state safety}~\cite{ames2019issf}.
For simplicity, we state this theorem only for the safety of multiplicative compartments.

\begin{thm}
\label{theo:issf}
Consider dynamical system~(\ref{eq:closedloop_disturbance}), function $h$ in~(\ref{eq:h_multi}) and 
the corresponding set $\mathcal{S}$ given by~(\ref{eq:safeset}).
Assume that the nominal controller $u(t)$ guarantees safety without the input disturbance $d(t)$ by satisfying~(\ref{eq:safetycondition_multi}), while the input disturbance $d(t)$ defined by~(\ref{eq:disturbance}) is bounded by ${\| d(t) \|_{\infty} \leq \delta}$.
Then, set $S$ is input-to-state safe in the sense that a larger set ${\mathcal{S}_{d} \supseteq S}$ given by
\begin{equation}
\mathcal{S}_{d} := \{ x  \in \mathbb{R}^{n+m} ~ : ~ h_{d}(x) \geq 0 \}
\end{equation}
is forward invariant (safe) under dynamics~(\ref{eq:closedloop_disturbance}), where ${h_{d}: \mathbb{R}^{n+m} \to \mathbb{R}}$ is defined by
\begin{equation}
h_{d}(x) := h(x) + \frac{\delta}{\alpha} \| g_{i}(w(t)) \|_{\infty}.
\label{eq:hd}
\end{equation}
\end{thm}

\proof
Similarly to~(\ref{eq:safetycondition}) and~(\ref{eq:safetycondition_extended}), the necessary and sufficient condition for the invariance of $\mathcal{S}_{d}$ is given by
\begin{equation}
\dot{h}_{d}(x(t))
\geq - \alpha h_{d}(x(t)).
\end{equation}
Substituting~(\ref{eq:hd}), using ${\dot{h}_d(x(t)) = \dot{h}(x(t))}$, and taking the derivative along the solution of~(\ref{eq:closedloop_disturbance}) yields
\begin{equation}
- \varphi_{i}(w(t)) - g_{i}(w(t)) (u(t) \!+\! d(t)) \!+\! \delta \| g_{i}(w(t)) \|_{\infty} \!\geq 0,
\end{equation}
which indeed holds, since~(\ref{eq:safetycondition_multi}) and
${\| d(t) \|_{\infty} \leq \delta}$
hold.
\hfill \qedsymbol

How much larger set $\mathcal{S}_{d}$ is compared to set $\mathcal{S}$ depends on the size $\delta$ of the disturbance that is related to the prediction error $\varepsilon$.
If the prediction is perfect (${x_{\rm p}(t) = x(t)}$), then ${\varepsilon=0}$, ${\delta = 0}$ and $\mathcal{S}_{d}$ recovers $\mathcal{S}$.
However, if one implements a delayed state feedback controller without prediction (${x_{\rm p}(t) = x(t-\tau)}$), then $\varepsilon$ and $\delta$ can be large, while $\mathcal{S}_{d}$ can be significantly larger than the desired set $\mathcal{S}$.

\begin{exmp}
\label{exmp:pred}
A possible model-based prediction can be done as follows.
At each time moment $t$, we take the most recent available measurement ${x(t-\tau)}$ and calculate the predicted state $x_{\rm p}(t)$ by numerically integrating the ideal delay-free closed loop over the delay interval ${\theta \in [t-\tau,t]}$:
\begin{align}
\begin{split}
\dot{w}_{\rm p}(\theta) & = f(w_{\rm p}(\theta)) + g(w_{\rm p}(\theta)) A(x_{\rm p}(\theta)), \\
\dot{z}_{\rm p}(\theta) & = q(w_{\rm p}(\theta)) + r(z_{\rm p}(\theta)),
\end{split}
\label{eq:predictor}
\end{align}
where ${x_{\rm p} = [w_{\rm p}^{\rm T}\ z_{\rm p}^{\rm T}]^{\rm T}}$ and the initial condition for integration is ${x_{\rm p}(t-\tau) = x(t-\tau)}$.
The thick red curves in Figs.~\ref{fig:SIRcontrol} and~\ref{fig:SIHRDcontrol} involve this kind of predictor feedback to compensate the delay $\tau$ in an ideal scenario with accurate predictor.
The thin red lines in Fig.~\ref{fig:SIHRDcontrol} illustrate the effect of prediction error: when ${\tau \pm 0.33 \tau}$ are used during prediction instead of $\tau$, the strict safety guarantees may be lost, but a level of input-to-state safety is ensured by keeping a larger set safe.
\end{exmp}

\section{CONCLUSIONS}
\label{sec:concl}

We viewed compartmental epidemiological models as control systems where human actions (such as quarantining or social distancing) are considered as control input.
By the framework of control barrier functions, we synthesized optimal safety-critical active intervention controllers that formally guarantee safety against the spread of infection while keeping mitigation efforts minimal. 
We highlighted that time delays arising during state measurements can significantly affect safety-critical control, and we proposed predictor feedback to compensate the delays while preserving a certain level of input-to-state safety.
We demonstrated our results on compartmental models fitted to US COVID-19 data, where we synthesized controllers to keep infection, hospitalization and deaths within prescribed limits.
Although translating the proposed continuous control actions into discrete policies is nontrivial, the controllers can guide policy makers to decide whether mitigation efforts shall be reduced or increased, and this recommendation can be ever updated based on new data.





\section*{ACKNOWLEDGMENT}

The authors would like to thank Franca Hoffmann and G\'abor St\'ep\'an for the valuable discussions on this topic.
This research is supported in part by the National Science Foundation, CPS Award \#1932091.





\bibliographystyle{IEEEtran}
\bibliography{covidref}

\end{document}